\documentclass[runningheads]{llncs}

\usepackage{xspace}
\usepackage{mathrsfs}
\usepackage{amsmath}
\usepackage{amssymb}
\usepackage{amsfonts}
\usepackage{times}
\usepackage{graphicx}
\usepackage{latexsym}
\usepackage{bm}
\usepackage{float}
\usepackage{url}

\makeatletter{}\usepackage{import}

\usepackage{ifthen}
\usepackage{url}
\usepackage{amssymb}
\usepackage{amsmath}
\usepackage{import}
\usepackage{xparse}
\usepackage{amsmath}

\usepackage[usenames,dvipsnames]{xcolor}
\usepackage{xspace}
\usepackage{datatool}
\usepackage{glossaries}

\newcommand{\m}[1]{\ensuremath{#1}\xspace}

	\newcommand{\lrule}{\m{\leftarrow}}
	\newcommand{\cause}{\m{\stackrel{c}{\lrule}}}

	\newcommand{\struct}{\m{I}}

	\NewDocumentCommand\inter{g+g}{%
	  \IfNoValueTF{#1}
	    {\struct}
	    {\m{#1^{#2}}}}

	\renewcommand{\int}{\m{\mathbb{Z}}}

	\NewDocumentCommand\subs{g+g}{%
	  \IfNoValueTF{#1}
	    {\m{/}}
	    {\m{#1/ #2}}}

	\newcommand{\logicname}[1]{\text{\sc #1}\xspace}

	\newcommand{\idp}{\logicname{IDP}}

	\newcommand{\minisatid}{\logicname{MiniSAT(ID)}}

	\newcommand{\fodotidp}{\logicname{FO(\ensuremath{\cdot})\ensuremath{^{\mathtt{IDP}}}}}
	
	\newcommand{\fodot}{\logicname{FO(\ensuremath{\cdot})}}

\newcommand{\ouracronym}[3]{%
	\newacronym{#1}{#2}{#3}
	\expandafter\newcommand\csname #1\endcsname{\gls{#1}\xspace}%
}
	\ouracronym{FO}{FO}{first-order logic}
	\ouracronym{MX}{MX}{Model Expansion}
	\ouracronym{MO}{MO}{Model Optimization}
	\ouracronym{ASP}{ASP}{Answer Set Programming}
	\ouracronym{TP}{TP}{Theorem Proving}
	\ouracronym{LP}{LP}{Logic Programming}
	\ouracronym{CP}{CP}{Constraint Programming}
	\ouracronym{KR}{KR}{Knowledge Representation}
	\ouracronym{CSP}{CSP}{Constraint Satisfaction Problem}
	\ouracronym{SMT}{SMT}{SAT Modulo Theories}
	\ouracronym{KBS}{KBS}{knowledge-base system}
	\ouracronym{NNF}{NNF}{Negation Normal Form}
	\ouracronym{FNNF}{FNNF}{Flat Negation Normal Form}
	\ouracronym{DefNNF}{DefNNF}{Definition Negation Normal Form}
	\ouracronym{CDCL}{CDCL}{Conflict-Driven Clause-Learning}
	\ouracronym{LCG}{LCG}{Lazy Clause Generation}

	\makeatletter
	\def\ifenv#1{
	\def\@tempa{#1}%
	\def\@ttempa{#1*}%
	\ifx\@tempa\@currenvir
	\expandafter\@firstoftwo
	\else
	\expandafter\@secondoftwo
	\fi
	}
	\makeatother

	\newcommand{\ddrule}[4]{\ensuremath{#1 \leftarrow #2 & \{#3\} & #4}}
	\newcommand{\drule}[2]{\ensuremath{#1 & \leftarrow & #2}}

	\newcommand{\darule}[4]{\ensuremath{#1 \leftarrow #2 & \{#3\} & #4}}
	\newcommand{\arule}[2]{\ensuremath{#1 \, &\leftarrow \, #2}}

	\newcommand{\LNDRule}[2]{
	\ifenv{array}
	{\drule{#1}{#2}}
	{ \ifenv{align}
		{\arule{#1}{#2}}
		{\ifenv{align*}
		{\arule{#1}{#2}}
		{ERROR: using LDRule in unsupported environment: \@currenvir}
		}
	}
	}

	\newcommand{\LDRule}[4]{
	\ifenv{array}
	{\ddrule{#1}{#2}{#3}{#4}}
	{ \ifenv{align}
		{\darule{#1}{#2}{#3}{#4}}
		{\ifenv{align*}
		{\darule{#1}{#2}{#3}{#4}}
		{ERROR: using LDRule in unsupported environment: \@currenvir}
		}
	}
	}

	\NewDocumentCommand\LRule{m+g+g+g}{%
		\IfNoValueTF{#2}%
		{#1.&}{%
		\IfNoValueTF{#3}
		{\LNDRule{#1}{#2}}
		{\LDRule{#1}{#2}{#3}{#4}}%
		}
	}

	\NewDocumentCommand\CLRule{m+g}{%
	\ifenv{array}
	{\cdrule{#1}{#2}}
	{ \ifenv{align}
		{\carule{#1}{#2}}
		{\ifenv{align*}
			{\carule{#1}{#2}}
			{ERROR: using CLRule in unsupported environment: \@currenvir}
		}
	}
	}

	\NewDocumentCommand\carule{m+g}{%
		\IfNoValueTF{#2}
			{\ensuremath{#1.}}
			{\ensuremath{#1 \, &\cause \, #2}}}
	\NewDocumentCommand\cdrule{m+g}{%
		\IfNoValueTF{#2}
			{\ensuremath{#1.}}
			{\ensuremath{#1 & \cause & #2}}}

	\newcommand{\algrule}[4]{
	\hbox{{#1}:}& 
	\quad #2 ~\longrightarrow~ #3 
	\hbox{~ if } #4\\
	}

	\newcommand{\AlgoRule}[4]{
	\ifenv{array}
	{\algrule{#1}{#2}{#3}{#4}}
		{ERROR: using AlgoRule in unsupported environment: \@currenvir}
	}

\newcommand{\commentstyle}{\color{Gray}}

	\usepackage{listings}

	\lstdefinelanguage{idp}{
		morekeywords=[1]{namespace,vocabulary,theory,structure,procedure,term,set,formula, spec, specification},
		morekeywords=[2]{include,using,type,isa,contains,partial,extern,LFD,GFD,constructed,from,constraint,func,pred,supertype,of,subtype,define},
		morekeywords=[3]{int,float,char,string,nat},
		morekeywords=[4]{if,then,else,for,end},
		morecomment=[s]{/*}{*/},	
		morecomment=[l]{//}
	}
	\newcommand{\code}[1]{\texttt{#1}}

	\newcommand{\ignore}[1]{}

	\newboolean{nocomments}
	\setboolean{nocomments}{false}

	\newboolean{commentmargin}
	\setboolean{commentmargin}{true}

	\newcommand{\namedcomment}[3]{
		\ifthenelse{\boolean{nocomments}}
		{} 
		{ 
			\ifthenelse{\boolean{commentmargin}}
				{ {\color{#3} \marginpar{\color{#3}\sc #2}#1}  } 
				{  {\color{#3} {\sc #2}: #1}  } 
		}
	}
	\newcommand{\mnamedcomment}[3]{\ifthenelse{\boolean{nocomments}}{}{{\marginpar{ \color{#3}{\sc #2}:#1}}}}

	\usepackage{soul}

\newcommand{\keyword}[2]{%
	\expandafter\newcommand\csname #1\endcsname{#2\xspace}%
	\expandafter\newcommand\csname #1s\endcsname{#2s\xspace}%
	\expandafter\newcommand\csname #1ness\endcsname{#2ness\xspace}%
}
 
\usepackage{tikz}
\usetikzlibrary{arrows}
\usetikzlibrary{positioning}

\urldef{\mailsa}\path|{marcos.cramer,diego.ambrossio}@uni.lu|
\urldef{\mailsb}\path|{pieter.vanhertum,marc.denecker}@cs.kuleuven.be|
\newcommand{\keywords}[1]{\par\addvspace\baselineskip
\noindent\keywordname\enspace\ignorespaces#1}

\begin{document}

\mainmatter  
\title{Modelling Delegation and Revocation Schemes in IDP}

\author{Marcos Cramer, Pieter Van Hertum, Diego Agust\'{i}n Ambrossio, Marc Denecker}

\institute{University of Luxembourg\\
\mailsa\\
KU Leuven, Belgium\\
\mailsb}

\maketitle

\begin{abstract}
In ownership-based access control frameworks with the possibility of delegating permissions and administrative rights, chains of delegated accesses will form. There are different ways to treat these delegation chains when revoking rights, which give rise to different revocation schemes. In this paper, we show how IDP -- a knowledge base system that integrates technology from ASP, SAT and CP -- can be used to efficiently implement executable revocation schemes for an ownership-based access control system based on a declarative specification of their properties.
\end{abstract}

\section{Introduction}

In ownership-based frameworks for access control, it is common to allow principals (users or processes) to grant both permissions and administrative rights to other principals in the system. Often it is desirable to grant a principal the right to further grant permissions and administrative rights to other principals. This may lead to delegation chains starting at a \emph{source of authority} (for example the owner of a resource) and passing on certain permissions to other principals in the chain. 

Furthermore, such frameworks commonly allow a principal to revoke a permission that she granted to another principal. Depending on the reasons for the revocation, different ways to treat the chain of principals whose permissions depended on the second principal's delegation rights can be desirable. For example, if one is revoking a permission given to an employee because he is moving to another position in the company, it makes sense to keep in place the permissions of principals who received their permissions from this employee; but if one is revoking a permission from a user who has abused his rights and is hence distrusted by the user who granted the permission, it makes sense to delete the permissions of principals who received their permission from this user. Any algorithm that determines which permissions to keep intact in which permissions to delete in the case of the revocation of a permission is called a \emph{revocation scheme}.

Hagstr\"om et al. \cite{Hagstrom} have presented a framework for classifying possible revocation schemes along three different dimensions: the extent of the revocation to other grantees (propagation), the effect on other grants to the same grantee (dominance), and the permanence of the negation of rights (resilience). Since there are two options along each dimension, there are in total eight different revocation schemes in Hagstr\"om's framework. This classification was based on revocation schemes that had been implemented in database management systems \cite{Griffiths,Fagin,Bertino96,Bertino97}.

\idp is a \emph{Knowledge Base System}, which combines a declarative specification in \fodot, with imperative management of the specification via the Lua scripting language. An \fodot specification theory consists of formulas in first-order logic and \emph{inductive definitions}. 
Inductive definitions are essentially logic programs in which clause bodies can contain arbitrary first-order formulas.
The combination of the declarative specification and the imperative programming environment makes this logic programming tool suitable for solving a large variety of different problems.

In this paper, we show that revocation schemes can be efficiently implemented in \idp by modelling them as \idp theories. One of the key features that make \idp a very efficient tool for implementing revocation schemes is the possibility to use inductive definitions for defining functions and predicates in \fodot, since the formal definition of revocation schemes can be captured in an elegant way as an inductive definition. 

The paper is structured as follows: We introduce Hagstr\"om et al.'s classification of revocation schemes in section \ref{rs}. After introducing \fodot and \idp in section \ref{IDP}, we show how we implemented the revocation schemes of Hagstr\"om et al.'s classification in \idp in section \ref{prototyping}. Section \ref{related} discusses related work and section \ref{conclusion} concludes the paper.

\makeatletter{}\section{The revocation classification framework}
\label{rs}
In this section we give both a formal and an informal presentation of Hagstr\"om et al.'s \cite{Hagstrom} classification of revocation schemes. 

Let $P$ be the set of principals (users or processes) in the system, let $O$ be the set of objects for which authorizations can be stated and let $A$ be the set off access types, i.e.\ of actions that principals may perform on objects. For every object $o \in O$, there is a \emph{source of authority} (\emph{SOA}), for example the owner of file $o$, which is a principal that has full power over object $o$ and is the ultimate authority with respect to accesses to object $o$. For any $a \in A$ and $o \in O$, the SOA of $o$ can grant the right to access $a$ on object $o$ to other principals in the system, and can also delegate the right to grant access and to grant this delegation right. Additionally, our framework allows for negative authorizations, which can be used to temporarily block a principal's access or delegation rights concerning a certain object and access type. 

We assume that all authorizations in the system are stored in an authorization specification, and that every authorization is of the form $(i,j,a,o,b_1,b_2)$, where $i,j \in P$, $a \in A$, $o \in O$ and $b_1$ and $b_2$ are booleans. The meaning of this authorization is that principal $i$ is granting some permission concerning access type $a$ on object $o$ to principal $j$. If $b_1$ is $\top$, then the permission is a positive permission, else it is a negative permission. If $b_2$ is $\top$, the permission contains the right to delegate the permission further. Since it does not make sense to combine a negative permission with the right to delegate the permission, the combination $\bot,\top$ for $b_1,b_2$ is disallowed.

There is no interaction between the rights of principals concerning different access-object pairs $(a,o)$. For this reason, we can consider $a$ and $o$ to be fixed for the rest of the paper, and can simplify $(i,j,a,o,b_1,b_2)$ to $(i,j,b_1,b_2)$.

\subsection{Delegation chains and connectivity property}
\label{delegation}
We first present the part of the system that does not involve negative authorizations. In section \ref{negative} we will introduce negative authorizations.

The right of a principal $i$ to delegate the access right to other principals can be defined by the existence of a \emph{rooted delegation chain}, i.e.\ a delegation chain connecting the SOA with $i$:

\begin{definition}
 A \emph{rooted delegation chain} for principal $i$ is a chain $(p_1,\dots,p_n)$ of principals satisfying the following properties: 
 \begin{enumerate}
  \item $p_1$ is the source of authority.
  \item $p_n$ is $i$.
  \item For every integer $k$ with $1 \leq k < n$, the authorization $(p_k,p_{k+1},\top,\top)$ is in place.
 \end{enumerate}
\end{definition}

A principal $j$ has the access right if she has delegation right or if some principal with delegation right has granted her the access right, i.e.\ if there is a principal $i$ such that the authorization $(i,j,\top,\bot)$ is in place and there is a rooted delegation chain for $i$.

The framework allows an authorization $(i,j,b_1,b_2)$ to be in the authorization specification only if $i$ has the delegation right. This is called the \emph{connectivity property}:

\vspace{4mm}
\noindent \textbf{Connectivity property:} 
 \textit{For every authorization $(i,j,b_1,b_2)$ in the authorization specification, there is a rooted delegation chain for $i$.}
\vspace{4mm}

We visualize an authorization specification by a labelled directed graph as in the following example:

\vspace{-3mm}
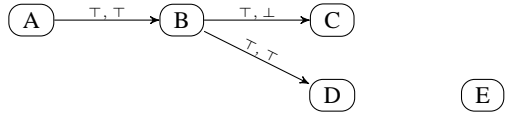
\begin{figure}[H]
\center
\begin{tikzpicture}[->,>=stealth',auto,node distance=4cm and 4cm,
                    main node/.style={rectangle,rounded corners=5,draw}]

  \node[main node] (A) at (0,1) {~A~~};
  \node[main node] (B) at (2,1) {~B~~};
  \node[main node] (C) at (4,1) {~C~~};
  \node[main node] (D) at (4,0) {~D~~};
  \node[main node] (E) at (6,0) {~E~~};

  \path[every node/.style={font=\sffamily\tiny},pos=0.5, sloped]
    (A) edge node [above=-.1] {$\top, \top$} (B)

    (B) edge node [above=-.1] {$\top, \bot$} (C)

    (B) edge node [above=-.1] {$\top, \top$} (D);

\end{tikzpicture}
\caption{Authorization specification visualized as labelled directed graph}
\end{figure}
\vspace{-3mm}
In this example, in which $A$ is the SOA (as in all forthcoming examples), the principals $A,B$ and $D$ have the delegation right, $C$ has the access right but not the delegation right, and $E$ has no rights concerning the access type and object in question.

\subsection{Negative authorizations}
\label{negative}

A negative authorization from $i$ to $j$ can inactivate a positive authorization from $i$ to $j$ without deleting it. The purpose of this is to make it possible to temporarily take away rights from a user without deleting anything from the authorization specification, so that it is easier to go back to the state that was in place before this temporary removal of rights. 

Hagstr\"om et al. \cite{Hagstrom} leave it open whether negative permissions dominate positive ones or the other way round. In this paper, we work under the assumption that positive permission dominate negative permissions.
More precisely, this means that a negative authorization $(i,j,\bot,\bot)$ directly inactivates only positive authorizations from $i$ to $j$, and leaves other permission to $j$ active. But the connectivity property is assumed to also hold for the subset of the authorization specification that consists only of the active authorizations. Hence additionally to the directly inactivated authorizations, there may also be indirectly inactivated authorizations, as the authorization from $B$ to $C$ in the following example:
\vspace{-3mm}
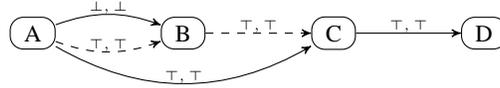
\begin{figure}[H]
\center
\begin{tikzpicture}[->,>=stealth',auto,node distance=4cm and 4cm,
                    main node/.style={rectangle,rounded corners=5,draw}]

  \node[main node] (A) at (0,1) {~A~~};
  \node[main node] (B) at (2,1) {~B~~};
  \node[main node] (C) at (4,1) {~C~~};
  \node[main node] (D) at (6,1) {~D~~};

  \path[every node/.style={font=\sffamily\tiny},pos=0.5, sloped]
    (A) edge [bend left=20] node [above=-.1] {$\bot, \bot$} (B)
        edge [dashed, bend right=20] node [above=-.1] {$\top, \top$} (B)
        edge [bend right=30] node [above=-.1] {$\top, \top$} (C)

    (B) edge [dashed] node [above=-.1] {$\top, \top$} (C)

    (C) edge node [above=-.1] {$\top, \top$} (D);

\end{tikzpicture}
\caption{The effect of a negative authorization}
\end{figure}
\vspace{-3mm}
In this example, principal $A$ has issued a negative authorization to principal $B$, thus inactivating the access and delegation rights of $B$. Since $B$ no longer has the right to delegate, the authorization from $B$ to $C$, which could only be issued because of $B$'s right to delegate, is also inactivated. But $C$ still has a rooted delegation chain that is independent of $B$, so that the authorization from $B$ to $C$ is not affected. 

In order to formally specify which authorizations get inactivated in this way, we first need to define the notion of an \emph{active rooted delegation chain}:
\begin{definition}
 An \emph{active rooted delegation chain} for principal $i$ is a chain $(p_1,\dots,p_n)$ of principals satisfying the following properties: 
 \begin{enumerate}
  \item $p_1$ is the source of authority.
  \item $p_n=i$.
  \item For every integer $k$ with $1 \leq k \leq n$, the authorization $(p_k,p_{k+1},\top,\top)$ is in place and the authorization $(p_k,p_{k+1},\bot,\bot)$ is not in place.
 \end{enumerate}
\end{definition}
Now a positive authorization $(i,j,\top,b)$ is considered inactive if there is no active rooted delegation chain for $i$. 

When the authorization specification contains negative authorizations, the delegation right and access right definitions of section \ref{delegation} can no longer be applied as stated before, but must be modified by adding the word ``active'' to ``rooted delegation chain'': A principal $i$ has delegation right if there is an \emph{active} rooted delegation chain for $i$, and a principal $j$ has access right if there is a principal $i$ such that the authorization $(i,j,\top,\bot)$ is in place and there is an \emph{active} rooted delegation chain for $i$.

\subsection{The three dimensions}
Hagstr\"om et al. \cite{Hagstrom} have introduced three dimensions according to which revocation schemes can be classified. These are called \emph{propagation}, \emph{dominance} and \emph{resilience}:

\subsubsection{Propagation.} 
The decision of a principal $i$ to revoke an authorization previously granted to a principal $j$ may either be intended to affect only the direct recipient $j$ or to affect all the other users in turn authorized by $j$. In the first case, we say that the revocation is \emph{local}, in the second case that it is \emph{global}. 

\subsubsection{Dominance.} 
This dimension deals with the case when a principal losing a permission in a revocation still has permissions from other grantors. 
If these other grantors' are dependent on the revoker, she can dominate these grantors and revoke the permissions from them. This is called a \emph{strong} revocation. The revoker can also choose to make a \emph{weak} revocation, where permissions from other grantors to a principal losing a permission are kept. 

In order to formalize this dimension, we need to define what we mean by a principal's delegation rights to be independent of another principal:
\begin{definition}
 A principal $j$ \emph{has delegation rights independent of} a principal $i$ iff there is an active rooted delegation chain $(p_1,\dots,p_n)$ such that $p_1$ is the SOA, $p_n=j$ and $p_k \neq i$ for every $1 \leq k \leq n$.
\end{definition}

\subsubsection{Resilience.} 
This dimension distinguishes revocation by removal of positive authorizations from revocation by negative authorizations which just inactivate positive authorizations. We call revocations of the first kind \emph{deletes} and revocations of the second kind \emph{negatives}.

\subsection{The eight revocation schemes}
For brevity, we just present five of the eight revocation schemes in detail, each with an example in which the authorization from $A$ to $B$ in the following authorisation specification is revoked according to the revocation scheme under consideration:

\vspace{-3mm}
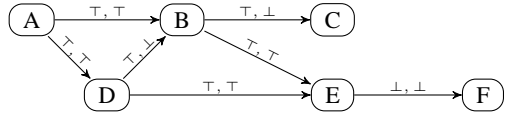
\begin{figure}[H]
\center
\begin{tikzpicture}[->,>=stealth',auto,
                    main node/.style={rectangle,rounded corners=5,draw}]

  \node[main node] (A) at (0,1) {~A~~};
  \node[main node] (B) at (2,1) {~B~~};
  \node[main node] (C) at (4,1) {~C~~};
  \node[main node] (D) at (1,0) {~D~~};
  \node[main node] (E) at (4,0) {~E~~};
  \node[main node] (F) at (6,0) {~F~~};

  \path[every node/.style={font=\sffamily\tiny},pos=0.5, sloped]
    (A) edge node [above=-.1] {$\top, \top$} (B)
        edge node [above=-.1] {$\top, \top$} (D)

    (B) edge node [above=-.1] {$\top, \bot$} (C)
        edge node [above=-.1] {$\top, \top$} (E)

    (D) edge node [above=-.1]  {$\top, \bot$} (B)
        edge node [above=-.1] {$\top, \top$} (E)

    (E) edge node [above=-.1] {$\bot, \bot$} (F);

\end{tikzpicture}
\caption{Example authorization specification before revocation}
\end{figure}
\vspace{-7mm}

\subsubsection{Weak local delete.}
A \emph{weak local delete} of a positive authorization from $i$ to $j$ has the following effect:
\vspace{-1mm}
\begin{enumerate}
 \item The authorization from $i$ to $j$ is deleted.
 \item If step 1 causes $j$ to lose its delegation right, all authorizations emerging from $j$ are deleted.
 \item For every authorization $(j,k,b_1,b_2)$ deleted in step 2, an authorization of the form $(i,k,b_1,b_2)$ is issued.
\end{enumerate}
\vspace{-1mm}
Step 2 ensures that the connectivity property is satisfied at $j$. This being a local revocation scheme, step 3 ensures that all rights that users other than $j$ had before the operation are intact.

\vspace{-3mm}
\begin{figure}[H]
\center
\begin{tikzpicture}[->,>=stealth',auto,node distance=4cm and 4cm,
                    main node/.style={rectangle,rounded corners=5,draw}]

  \node[main node] (A) at (0,1) {~A~~};
  \node[main node] (B) at (2,1) {~B~~};
  \node[main node] (C) at (4,1) {~C~~};
  \node[main node] (D) at (1,0) {~D~~};
  \node[main node] (E) at (4,0) {~E~~};
  \node[main node] (F) at (6,0) {~F~~};

  \path[every node/.style={font=\sffamily\tiny},pos=0.5, sloped]
    (A) edge [bend left=20, in=150, out=30, above=-.1] node {$\top, \bot$} (C)
        edge node [above=-.1] {$\top, \top$} (D)
        edge [bend right, in=215,out=270] node [below] {$\top, \top$} (E)

    (D) edge node [above=-.1] {$\top, \bot$} (B)
        edge node [above=-.1] {$\top, \top$} (E)

    (E) edge node [above=-.1] {$\bot, \bot$} (F);

\end{tikzpicture}
\vspace{-1mm}
\caption{Weak Local Delete from $A$ to $B$}
\end{figure}
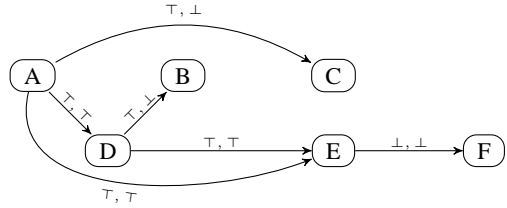
\vspace{-7mm}

\subsubsection{Weak global delete.}
A \emph{weak global delete} of a positive authorization from $i$ to $j$ has the following effect:
\begin{enumerate}
 \item The authorization from $i$ to $j$ is deleted.
 \item Recursively, any authorization emerging from a principal who loses her delegation right in step 1 or step 2 is deleted.
\end{enumerate}
The recursive step 2 ensures that the connectivity property is satisfied for the whole authorization specification after this operation.

\vspace{-3mm}
\begin{figure}[H]
\center
\begin{tikzpicture}[->,>=stealth',auto,node distance=4cm and 4cm,
                    main node/.style={rectangle,rounded corners=5,draw}]

  \node[main node] (A) at (0,1) {~A~~};
  \node[main node] (B) at (2,1) {~B~~};
  \node[main node] (C) at (4,1) {~C~~};
  \node[main node] (D) at (1,0) {~D~~};
  \node[main node] (E) at (4,0) {~E~~};
  \node[main node] (F) at (6,0) {~F~~};

  \path[every node/.style={font=\sffamily\tiny},pos=0.5, sloped]
    (A) edge node [above=-.1] {$\top, \top$} (D)

    (D) edge node [above=-.1] {$\top, \bot$} (B)
        edge node [above=-.1] {$\top, \top$} (E)

    (E) edge node [above=-.1] {$\bot, \bot$} (F);

\end{tikzpicture}
\caption{Weak Global Delete from $A$ to $B$}
\end{figure}
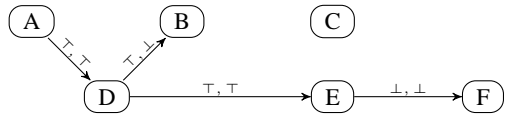
\vspace{-7mm}

\subsubsection{Strong local delete.}
A \emph{strong local delete} of a positive authorization from $i$ to $j$ has the following effect:
\vspace{-1mm}
\begin{enumerate}
 \item The authorization from $i$ to $j$ is deleted.
 \item Every authorization of the form $(k,j,\top,b)$ such that $k$ is not independent of $i$ is deleted.
 \item If steps 1 and 2 cause $j$ to lose its delegation right, all authorizations emerging from $j$ are deleted.
 \item For every authorization $(j,k,b_1,b_2)$ deleted in step 3, an authorization of the form $(i,k,b_1,b_2)$ is issued.
\end{enumerate}
\vspace{-1mm}
The only difference to the weak local delete is step 2, which is the step that makes this a strong revocation scheme.

\vspace{-3mm}
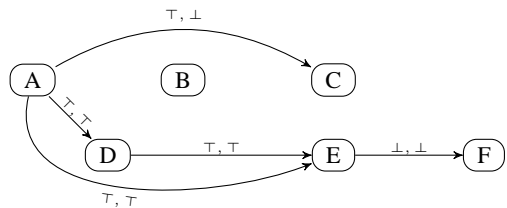
\begin{figure}[H]
\center
\begin{tikzpicture}[->,>=stealth',auto,node distance=4cm and 4cm,
                    main node/.style={rectangle,rounded corners=5,draw}]

  \node[main node] (A) at (0,1) {~A~~};
  \node[main node] (B) at (2,1) {~B~~};
  \node[main node] (C) at (4,1) {~C~~};
  \node[main node] (D) at (1,0) {~D~~};
  \node[main node] (E) at (4,0) {~E~~};
  \node[main node] (F) at (6,0) {~F~~};

  \path[every node/.style={font=\sffamily\tiny},pos=0.5, sloped]
    (A) edge [bend left=20, in=150, out=30] node {$\top, \bot$} (C)
        edge node [above=-.1] {$\top, \top$} (D)
        edge [bend right, in=215,out=270] node [below] {$\top, \top$} (E)

    (D) edge node [above=-.1] {$\top, \top$} (E)

    (E) edge node [above=-.1] {$\bot, \bot$} (F);

\end{tikzpicture}
\vspace{-1mm}
\caption{Strong Local Delete from $A$ to $B$}
\end{figure}
\vspace{-7mm}

\subsubsection{Strong global delete.}
A \emph{strong global delete} of a positive authorization from $i$ to $j$ has the following effect:
\vspace{-1mm}
\begin{enumerate}
 \item The authorization from $i$ to $j$ is deleted.
 \item Recursively, delete authorizations as follows:
 \begin{enumerate}
  \item Any authorization emerging from a principal who loses her delegation right in step 1, step 2.(a) or step 2.(b) is deleted.
  \item Any authorization of the form $(k,l,\top,b)$, where $l$ is a principal who loses her delegation right in step 1, step 2.(a) or step 2.(b) and $k$ is not independent of $i$, is deleted.
 \end{enumerate}
\end{enumerate}
\vspace{-1mm}
Here the recursive deletion procedure contains two different kinds of deletions: 2.(a) makes it a global revocation scheme and 2.(b) makes is a strong revocation scheme.

\vspace{-3mm}
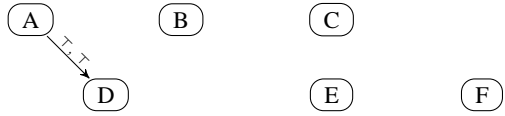
\begin{figure}[H]
\center
\begin{tikzpicture}[->,>=stealth',auto,node distance=4cm and 4cm,
                    main node/.style={rectangle,rounded corners=5,draw}]

  \node[main node] (A) at (0,1) {~A~~};
  \node[main node] (B) at (2,1) {~B~~};
  \node[main node] (C) at (4,1) {~C~~};
  \node[main node] (D) at (1,0) {~D~~};
  \node[main node] (E) at (4,0) {~E~~};
  \node[main node] (F) at (6,0) {~F~~};

  \path[every node/.style={font=\sffamily\tiny},pos=0.5, sloped]
    (A) edge node [above=-.1] {$\top, \top$} (D);

\end{tikzpicture}
\caption{Strong Global Delete from $A$ to $B$}
\end{figure}
\vspace{-7mm}

\subsubsection{Negative revocations.}
The negative revocations are similar to the positive revocations, only that instead of deleting positive authorizations, we inactivate them by issuing negative authorizations. We show this on the example of the weak global negative. The other three negative revocation schemes are adapted versions of the corresponding deletes in a similar way. 

A \emph{weak local negative} of a positive authorization from $i$ to $j$ has the following effect:\vspace{-1mm}\begin{enumerate}
 \item The negative authorization $(i,j,\bot,\bot)$ is added to the authorization specification.
 \item For every authorization $(j,k,b_1,b_2)$ inactivated by step 1, an authorization of the form $(i,k,b_1,b_2)$ is issued.
\end{enumerate}
\vspace{-1mm}
Unlike in the weak local delete, we do not delete any authorizations. The definition of \emph{inactive} authorizations from section \ref{negative} ensures that authorizations that get deleted in a weak local delete get inactivated in a weak local negative, even though we do not need to state explicitly which authorizations get inactivated.
\vspace{-3mm}
\begin{figure}[H]
\center
\begin{tikzpicture}[->,>=stealth',auto,node distance=4cm and 4cm,
                    main node/.style={rectangle,rounded corners=5,draw}]

  \node[main node] (A) at (0,1) {~A~~};
  \node[main node] (B) at (2,1) {~B~~};
  \node[main node] (C) at (4,1) {~C~~};
  \node[main node] (D) at (1,0) {~D~~};
  \node[main node] (E) at (4,0) {~E~~};
  \node[main node] (F) at (6,0) {~F~~};

  \path[every node/.style={font=\sffamily\tiny},pos=0.5, sloped]
    (A) edge [bend left=20, in=160, out=45] node [above=-.1] {$\top, \bot$} (C)
        edge [bend left=20] node [above=-.1] {$\bot, \bot$} (B)
        edge [dashed,bend right=10] node [above=-.1] {$\top, \top$} (B)
        edge  node [below=-.1] [above=-.1] {$\top, \top$} (D)
        edge [bend right, in=215,out=270] node [below] {$\top, \top$} (E)

    (B) edge [dashed] node [above=-.1] {$\top, \bot$} (C)
        edge [dashed] node [above=-.1] {$\top, \top$} (E)

    (D) edge node [above=-.1]  {$\top, \bot$} (B)
        edge node [above=-.1] {$\top, \top$} (E)

    (E) edge node [above=-.1] {$\bot, \bot$} (F);

\end{tikzpicture}
\caption{Weak Local Negative from $A$ to $B$}
\end{figure}
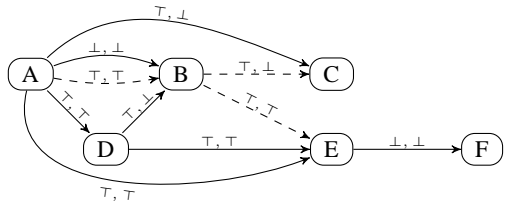
\vspace{-5mm}

\subsection{Undoing negative revocation schemes}
The purpose of negative authorizations is to temporarily block someone's rights. There is an \emph{undo operation}, which undoes the effect of such a temporary blocking.
In the case of the weak global negative, the undo operation just consists of deleting the negative authorization $(i,j,\bot,\bot)$ that was added to the authorization specification. In the case of the other three negative revocation schemes, we also need to delete the auxiliary authorizations that were issued as part of the revocation scheme. This can be achieved by labelling their dependence on the negative authorization $(i,j,\bot,\bot)$ that was added in step 1 of the negative revocation scheme, and deleting them as soon as this negative revocation is deleted. 

\makeatletter{}\section{\fodot and the \idp-system}
\label{IDP}
In developing the \fodot-\idp-project, the ambition is to create a full integration of existing \emph{pure declarative KR-languages} with a \emph{knowledge base system} to implement reasoning for these languages.
\subsection{The KR-languages: \fodot}
We use the term  \fodot for a family of extensions of first-order logic (FO). \fodot is developed with the purpose to combine ideas from multiple domains of knowledge representation, logic programming and on monotonic reasoning in a conceptual clear manner. The basis of this family of languages lies in first order (classical) logic, extended with new language constructs from the fields of logic programming, constraint programming and non monotonic reasoning.

The \idp-system supports an \fodot language, denoted by \fodotidp. In the context of this paper, the focus lies on \fodotidp, since this is the language used for the modelling of the revocation schemes further on. 
This section provides an overview of the core of the \fodotidp language, more details about the language can be found in \cite{WarrenBook/DeCatBBD14}. 
We assume familiarity with basic concepts of classical logic and logic programming.
The most important extensions in \fodotidp are \emph{types, arithmetic, (partial) functions, aggregates} and \emph{inductive definitions}. 

An \fodot specification consists of vocabularies, theories, terms and structures.
A vocabulary$\Sigma$ consists of a set of types and a set of predicate- and function symbols, annotated with the types of their arguments. 
A structure $\mathcal{S}$ over a vocabulary $\Sigma$ consists of a domain $D_T$, for every type $T$ and an interpretation for a subset of the other elements of $\Sigma$. 
A theory  $\mathcal{T}$ over a vocabulary $\Sigma$ consists of a set of FO sentences and a set of inductive definitions. 
Inductive definitions $\Delta$ are sets of rules of the form $\forall \overline{x}:P(\overline{x}) \leftarrow \varphi(\overline{y})$, with $\overline{y} \subset \overline{x}$ and $\varphi(\overline{y})$ a \fodotidp formula. 
We call predicate $P$ the defined predicate, or the head of the definition. 
Any other predicate or function symbol in $\Delta$ is called a parameter. 
The semantics used for these definitions are the well-founded semantics, as argued in \cite{tocl/DeneckerT08} this captures the intended meaning of all common forms of definitions  and extends the least model semantics of Prolog for negations. Informally a structure $\mathcal{S}$ satisfies $\Delta$ if the interpretation of a defined predicate $P$ in the well-founded model of $\mathcal{S}$, constructed relative to the restriction of $\mathcal{S}$ to the parameters of $\Delta$ is exactly the relation $P^{\mathcal{S}}$. 
Another extension in \fodotidp are the aggregates, these are functions over a set of domain elements, which map such a set to the sum, minimum, maximum, cardinality or product of the elements in that set.

A special property of the \fodot family of languages is that they are a ``true'' declarative logic. They can be used to create a specification of knowledge, not to formulate the description of a problem. Hence, it has no operational semantics and has no unique form of inference. This however does not mean we have no interest in solving problems with these logics. On the contrary, the disengagement of knowledge from problem solving and inferences makes that a specification of domain knowledge may be reused for solving multiple tasks and  problems \cite{iclp/DeneckerV08}. 
This leads to the idea of \idp, a knowledge base system, that manages a specification, and provides multiple inferences on it, in order to solve a whole range of problems using a single logic specification.

\subsection{The knowledge base system : \idp}
We define a knowledge base system (KBS) as a system that supports multiple inferences that can be used on a single specification to be able to execute a range of tasks.
The \idp (\emph{Imperative Declarative Programming}) framework is such a KBS, it combines a declarative specification
  \footnote{We use \idp syntax in the examples throughout the paper.
    Each \idp operator has an associated logical operator, the main (non-obvious) operators being: \code{\&}($\land$), \code{|}($\lor$), $\sim$($\lnot$), \code{!}($\forall$), \code{?}($\exists$), $<=>$($\equiv$), $\sim$\code{=}($\neq$).
  }
, in \fodotidp, a set of inferences and an imperative manipulation of the specification via the Lua \cite{SPE/IerusalimschyFC96} imperative programming environment \cite{inap/DePooterWD11}.
We will focus only on the most important inferences, for others and more details, we refer to \cite{WarrenBook/DeCatBBD14}.
\begin{description}
 \item[Modelchecking] Given a structure $I$, a theory $T$, modelchecking outputs true iff $I\models T$
 \item[Model expansion (mx)] Given a vocabulary $V$, a \fodot theory $T$, and a three-valued structure $I$, that contains a domain $D_T$ for every type in $T$, model expansion outputs two-valued structures $I'$ such that every $I'\models T$.
 \item[Propagation] Given a \fodot theory $T$ and a three-valued structure $I$, propagation returns a new three-valued structure $I'$ such that $I'$ approximates every model of $T$ and is more precise then $I$.
 \item[Deduction] Given a \fodot theory $T$ and a FO sentence $\varphi$, deduction outputs true iff $T\models \varphi$. Note: this method is sound but incomplete.
 \item[Progression] 
 In~\cite{iclp/BogaertsTemp}, LTC-theories (Linear Time Calculus) are proposed, a syntactic subclass of \fodot theories that allow to naturally model dynamic systems. 
 Given an LTC theory and a structure $I_n$ that provides information about the state of the system on a time point $n$, the \emph{progression} inference can be used to compute the state (or the possible states) at time point n+1 as a new structure $I_{n+1}$. 
 Repeating the process, we can compute all subsequent states, effectively simulating the dynamic system defined by T. 
 An LTC-theory consists of 3 types of constraints: constraints about the initial situation ($P(0)$), invariants $(\forall t: P(t) \lor Q(t))$, and "bistate" formula's that relates the state on the current point in time with that of the next $(\forall t: P(t) \Rightarrow P(t+1))$.
\ignore{ A small example is included in listing \ref{progress}, which also contains structure $I_1$ the result of the progression.
 \begin{lstlisting}[caption={A dynamic system: a lamp.},label={progress}]
vocabulary V {
    type Time
    LampOn(Time)
    PushSwitch(Time)
}
theory sp_theory1: sp_voc {
    //Constraints about the initial state
    ~LampOn(0).
    //BiState constraints
    !t:~LampOn(t)&PushSwitch(t)=>LampOn(t+1).
    !t:LampOn(t)&PushSwitch(t)=>~LampOn(t+1).
}
structure I_0 {
    Time=0..5
    LampOn={}
    PushSwitch={0}
}
structure I_1 {
    Time=0..5
    LampOn={1}
    PushSwitch={}
}
\end{lstlisting}}
 
 \end{description}
 The imperative programming environment supports a rich set of operators and inferences to take the logical building blocks in an \fodotidp specification (vocabularies, theories, terms and structures) and use these to manipulate them and solve more complex reasoning tasks.
 
 The efficiency of the \idp framework in solving problems has been proved in different settings, like for example the most recent ASP-competitions~(\cite{ASP-Comp-2},~\cite{ASP-Comp-3},~\cite{conf/lpnmr/Alviano}) and in applications (~\cite{corr/Blockeel13}, ~\cite{iclp/VanHertum13}). Also the different parts of the system have proved their use in multiple situations: the search algorithm \minisatid has been demonstrated in~\cite{misc/amadini2013}, where it turned out to be the single-best solver in their MiniZinc portfolio, and in the latest MiniZinc challenges~\cite{url:MinizincChallenge2012}. Next to this \idp is used as a didactic tool in various logic-oriented courses, at KU Leuven and at the University of Luxembourg, among others because of its close adherence to \FO and its support for deduction.

We look at a specific small example, the connected graph problem (Listing~\ref{mincolor}), given a graph, we want to know if it is fully connected.
In our vocabulary, we have a type node (the domain of nodes in the graph), a predicate \texttt{edge(node,node)} (there is an edge between these two nodes) and a predicate \texttt{reaches(node,node)} (this expresses the reachability relation between two nodes).
The theory contains our definitions and constraints.
We have one (inductive) definition in this theory, which defines reaches (definitions are given between ``\{'' and ``\}'').
Next to this, the theory contains 1 constraint: Every 2 nodes should be reachable from one another.

Besides these 3 logical building blocks, we also have the procedural Lua part.
In this code the solver is called to check a model and print out if the graph is fully connected or not.

\begin{lstlisting}[caption={Calling main() solves the graph connectivity problem for the given data.},label={mincolor}]
vocabulary sp_voc {
    type node
    edge(node,node)
    reaches(node,node)
}
theory sp_theory1: sp_voc {
    { reaches(x,y) <- edge(x,y).
      reaches(x,y) <- ?z: reaches(x,z) & reaches(z,y).   }
    !x y: reaches(x,y).  
}
structure sp_struct: sp_voc {
    node = {A..D} // shorthand for A,B,C,D
    edge = {A,B; B,C; C,D; A,D} // `;' separated list of tuples
}
procedure main() {
    sol = modelexpand(sp_theory1,sp_struct,lengthOfPath)[1]
    if (sol == nil) //If no result is returned, no models exist
    then print("The graph is not fully connected.\n")
    else print("The graph is fully connected.\n")
    end
}
\end{lstlisting}
Note that the concept of inductive definitions is essential to express what reachable means. 
They capture the construction of the reachability relation. 
First we take all edges: if there is an edge from node $x$ to node $y$, then $y$ is reachable from $x$. 
When we have added all the edges to $reaches$, we start using these to recursively add new tuples to our relation.If we would try to capture this without definitions, we would soon notice that this is impossible. Say we use the material implication to model the reaches relation:  \\[-25pt]
\begin{align*} 
  & reaches(x,y) \Leftarrow edge(x,y).\\
  & reaches(x,y) \Leftarrow \exists z : reaches(x,z) \land reaches(z,y).
\end{align*}
The relation $edge$ can now be empty even though every pair of nodes is in $reaches$. If we use an equivalence, we get following formulation: \\[-15pt]
\begin{align*} 
  & reaches(x,y) \Leftrightarrow edge(x,y)  \lor \exists z :reaches(x,z) \land reaches(z,y).
\end{align*}
Still we can have a model in which $edge$ is empty and every possible tuple $(x,y)\in reaches^I$.

\makeatletter{}\section{Modelling the revocation schemes in IDP}
\label{prototyping}

Aucher et al. \cite{Aucher} presented a formalization of the eight revocation schemes introduced in section \ref{rs} in a dynamic variant of propositional logic that resembles imperative programming languages. In this section we sketch our implementation\footnote{The implementation can be downloaded at \url{http://icr.uni.lu/mcramer/index.php?id=3}.} of the eight revocation schemes and the undoing operation in IDP. Because of the nature of IDP, whose inductive definitions suit the recursive character of the revocation schemes very well, the revocation schemes could be implemented in a very straightforward way. The implementation sheds light on both the formal properties and the practical implications of the revocation schemes, and can thus support a developer of an access control system in her decisions concerning the precise nature of the revocation schemes to be included in the system.

Unlike the formal definition in \cite{Aucher}, our implementation does not work by implementing each of the eight revocation schemes separately, but by specifying the formal properties of the three dimensions of the classification in an IDP theory. IDP can then execute the revocation schemes based on this formal specification.

In the sketch of the IDP implementation, we concentrate on the four deletion schemes. The only additional complication in the four negative schemes is the labelling system for keeping track of what to do in the undoing operation.\footnote{This labelling system is well-documented in the comments to the code of our implementation.}

\subsection{Preliminaries: Vocabulary and auxiliary predicates}
The IDP theory models the change of the authorization specification over time. Principals are modelled as objects of the theory's domain, whereas authorizations are modelled by a partial function (for positive authorizations) and a predicate (for negative authorizations) on pairs of principals. The authorizations cannot be modelled as objects, because they change over time, while IDP assumes a constant domain of objects. 

The partial function \texttt{pos\_{}auth} that models positive authorization can take \texttt{TT} and \texttt{TF} as values, depending on whether it represents an authorization of the form $(i,j,\top,\top)$ or $(i,j,\top,\bot)$. Apart from the two principals $i$ and $j$, it takes a point in time as argument. Negative authorizations are modelled by a separate predicate called \texttt{FF}, also taking a point in time and two principals as arguments. The reason for this separation of positive and negative authorizations is that it does not make sense to have two different positive authorizations linking the same pair of principals, whereas it does make sense to have a negative authorization additionally to a positive authorization linking the same pair of principals.

The objects \texttt{TT} and \texttt{TF} that serve as values of \texttt{pos\_{}auth} are given a separate type called \texttt{authorization}. The other types of objects in the domain of the IDP theory are points in time, principals and revocation schemes.

\begin{lstlisting}[caption={The vocabulary of the IDP implementation},label={vocab}]
vocabulary V{
	type time isa int // Points in times are integers
	type principal
	type scheme
	type authorization
	
	SOA:principal
	TT:authorization
	TF:authorization
	WGD:scheme // Weak Global Delete
	WLD:scheme // Weak Local Delete
	SGD:scheme // Strong Global Delete
	SLD:scheme // Strong Local Delete
	WGN:scheme // Weak Global Negative
	WLN:scheme // Weak Local Negative
	SGN:scheme // Strong Global Negative
	SLN:scheme // Strong Local Negative
	UN:scheme // Undo operation
	
	// Positive and negative authorizations
	partial pos_auth(time,principal,principal):permission
	FF(time,principal,principal)
	// Start configuration of authorizations
	partial pos_auth_start(principal,principal):permission
	FF_start(principal,principal)
	
	// Auxiliary predicates
	active_chain(time,principal)
	ind(time,principal,principal)
	access_right(time,principal) 
	
	// Predicate for specifying which revocation schemes to apply
	rs(time,scheme,principal,principal)
	
	// Changes on the authorization specification
	delete(time,principal,principal) 
	partial new(time,principal,principal):permission
}
\end{lstlisting}
\texttt{pos\_{}auth} is defined inductively by setting its values at time $t=0$ to the start configuration specified by \texttt{pos\_{}auth\_{}start}, and by modifying its values between time $t$ and time $t+1$ according to the changes specified by \texttt{delete} and \texttt{new}:
\begin{lstlisting}[caption={The definition of the authorization specification at a given time},label={specification}]
{ pos_auth(0,p1,p2)=x<-pos_auth_start(p1,p2)=x.
  pos_auth(t+1,p1,p2)=x<-pos_auth(t,p1,p2)=x & ~delete(t,p1,p2).
  pos_auth(t+1,p1,p2)=x<-new(t,p1,p2)=x.}
\end{lstlisting}
Since in this sketch of the implementation we are leaving out the negative revocation schemes, we can ignore negative authorizations. In the actual implementation, there are predicates \texttt{FF\_{}delete} and \texttt{new\_{}FF} that specify changes on the negative authorizations, and \texttt{FF} is defined in a way analogous to \texttt{pos\_{}auth} using these change predicated instead of \texttt{delete} and \texttt{new}.

The auxiliary predicates \texttt{active\_{}chain}, \texttt{ind} and \texttt{access\_{}right} model the existence of an active rooted delegation chain for a principal, the independence of a principal from another principal and the access right of a principal. Their definitions in IDP correspond directly to the definitions of the corresponding notions in section \ref{rs}:
\begin{lstlisting}[caption={The definitions of the auxiliary predicates},label={auxiliary}]
{ active_chain(t,SOA).
  active_chain(t,p1) <- ?p2: active_chain(t,p2) & pos_auth(t,p2,p1)=TT & ~FF(t,p2,p1). }
{ ind(t,SOA,p).
  ind(t,p1,p2) <- ?p: ~p=p2 & ind(t,p,p2) & pos_auth(t,p,p1)=TT & ~FF(t,p,p1). }
{ access_right(t,p) <- active_chain(t,p).
  access_right(t,p) <- ?p1:active_chain(t,p1) & pos_auth(t,p1,p)=TF & ~FF(t,p1,p). }	
\end{lstlisting}

\subsection{Specifying propagation and dominance for deletion schemes in the IDP theory}
The meaning of \emph{local} vs. \emph{global} propagation is captured by the inductive definition of the partial function \texttt{new}, which specifies which new authorizations are added to the authorization specification:
\begin{lstlisting}[caption={The definition of \texttt{new} captures the propagation dimension},label={new}]
{ new(t,i,k)=x <- ?j s:(s=WLD | s=SLD | s=WLN | s=SLN) & rs(t,s,i,j) & ~active_chain(t+1,j) & pos_auth(t,j,k)=x.}
\end{lstlisting}
Informally, this definition says that if in a local revocation scheme revoking a positive authorization from principal $i$ to principal $j$, $j$ is losing its delegation right, then every positive authorization from $j$ to another principal $k$ must be replaced by a positive authorization of the same authorization type  from $i$ to $k$. 

In order to understand why this definition of \texttt{new} also ensures that the propagation of the deletion is blocked in local revocation schemes in the desired way, we need to look at the definition of \texttt{delete}. It is defined using an inductive definition with four clauses:
\vspace{-3mm}
\begin{lstlisting}[caption={The definition of \texttt{delete} captures the dominance dimension},label={delete}]
{delete(t,i,j)<-rs(t,s,i,j) & (s=WLD | s=WGD | s=SLD | s=SGD).
 delete(t,i,j)<-pos_auth(t,i,j)=x & ~active_chain(t+1,i).
 delete(t,k,j)<-rs(t,SLD,i,j) & pos_auth(t,k,j)=x & ~ind(t,k,i).
 delete(t,z,w)<-rs(t,SGD,i,j) & delete(t,p,w) & pos_auth(t,z,w)=x & ~ind(t,z,i).
\end{lstlisting}
Let us first concentrate on the first two clauses: The first clause just states that in any deletion revocation scheme from $i$ to $j$, the positive authorization from $i$ to $j$ is deleted. The second clause defines the propagation of deletion by specifying that any positive authorization from $i$ to $j$ gets deleted if $i$ is losing its delegation right. Since in local revocation schemes, the definition of \texttt{new} ensures that principals who had previously received their delegation right from $j$ will now receive it from $i$, the propagation gets blocked after $j$ in local revocation schemes, as desired.

The meaning of \emph{strong} vs. \emph{weak} dominance is captured by the third and fourth line of the inductive definition of \texttt{delete}: These lines specify the additional deletions that are needed in strong revocation schemes. 

Note that we needed to specify the additional strength of the deletion separately for strong local deletes and strong global deletes: This is because we wanted -- in line with the definition of the strong global delete in \cite{Hagstrom} and \cite{Aucher} -- a strong global delete from $i$ to $j$ not only to be strong in the sense of deleting other permissions to $j$ dependent on $i$, but also to delete other permissions dependent on $i$ to descendants of $j$. We doubt, however, whether this additionally strength of the strong global delete would actually be desirable in a real access control system: Strong revocation schemes are usually applied to distrusted principals, whose rights one wants to restrict as much as possible. But there is no reason why another principal, who has a rooted delegation chain independent of this distrusted principal, should have his rights removed only because he also has a rooted delegation chain dependent on the distrusted principal. The version of the strong global delete that 
we judge more reasonable is the one in which the fourth line of the inductive definition of \texttt{delete} is removed and the third line is also applied to the strong global delete.

This discussion of the details of the strength of the strong global delete illustrates how modelling revocation schemes in IDP can shed light on the properties of the revocation schemes in a way that can support a developer of an access control system in fixing the specification of the schemes to be implemented in the system. 

\makeatletter{}\section{Related Work}
\label{related}

The classification of revocation schemes used in this paper was first introduced by Hagstr\"om et al. \cite{Hagstrom}. Their paper, however, was rather informal in nature. 

The first formalization of this classification was presented by Aucher et al. \cite{Aucher}. They use a dynamic variant of propositional logic for their formalization. Unlike our specification of the revocation schemes in IDP, their formalization required all eight revocation schemes to be formalized separately.

Barker et al. \cite{Barker} have represented delegation-revocation models in terms of reactive Kripke models \cite{Gabbay}. They implement this approach by translating first-order representations of the reactive Kripke models into an equivalent Answer Set Programming form. Answer Set Programming is a logic programming approach close in nature to IDP.

The \idp system is maturing, as is shown by applications in multiple fields of interest.
In~\cite{corr/Blockeel13} a set of machine learning applications have been solved by an approach with the \idp system and the \fodot framework. 
Among others, this research showed a very elegant and efficient solution for the stemmatology application. 
Given different versions of an ancient text, the goal of stemmatology is to find which one is the original and which text is copied from which.
The problem was specified in a theory of one sentence and was able to solve large instances. 

Another application in which \idp was put to the test was in~\cite{iclp/VanHertum13}, in which a typical application from Business Rule Systems was taken and the behaviour was modelled in the \idp system. 
A comparison between the \idp and the Business Rule approach was made. 
The \idp system had some great advantages, like the possibility to reason in context of incomplete knowledge, the ability to reason hypothetically and in multiple directions.

\makeatletter{}\section{Conclusion}
\label{conclusion}

We have shown, how the knowledge base system IDP can be used for efficiently implementing the revocation schemes in Hagstr\"om et al.'s \cite{Hagstrom} classification. This implementation works by specifying the properties of the three dimensions of the classification in an IDP theory. By using the model expansion inference of IDP, this declarative specification becomes an executable program implementing the eight revocation scheme in the classification. We also illustrated how the IDP implementation can help to shed light on the formal properties of the revocation schemes and can thus support the development of ownership-based access control systems. 

\bibliographystyle{splncs03}
\bibliography{references,idp-latex/krrlib}

\end{document}